\documentclass[reprint,showpacs,amsmath,amssymb,aps]{revtex4-1}
\usepackage{graphicx}
\usepackage{dcolumn}
\usepackage[T1]{fontenc}
\usepackage{bm}
\usepackage[utf8x]{inputenc}
\usepackage[usenames]{xcolor}

\begin{document}
 \title {Modeling non-stationary, non-axisymmetric heat patterns in DIII-D tokamak}
 \author {D. Ciro$^1$, T. E. Evans$^2$, I. L. Caldas$^3$}
 \email{$^1$davidcirotaborda@usp.br\newline $^2$evans@fusion.gat.com\newline $^3$ibere@if.usp.br}
 \affiliation {$^{1,3}$ Department of Applied Physics, S\~ao Paulo University, S\~ao Paulo, CEP 05508-090, Brazil\\
 $^2$General Atomics, PO Box 85608, San Diego, CA 92186-5608, USA.}
 
\begin {abstract}
  Non-axisymmetric stationary magnetic perturbations lead to the formation of homoclinic tangles near the divertor magnetic saddle in tokamak discharges. These tangles intersect the divertor plates in static helical structures that delimit the regions reached by open magnetic field lines reaching the plasma column and leading the charged particles to the strike surfaces by parallel transport. In this article we introduce a non-axisymmetric rotating magnetic perturbation to model the time evolution of the three-dimensional magnetic field of a single-null {DIII-D} tokamak discharge developing a rotating tearing mode. The non-axiymmetric field is modeled using the magnetic signals to adjust the phases and currents of a set of internal filamentary currents that approximate the magnetic field in the plasma edge region. The stable and unstable manifolds of the asymmetric magnetic saddle are obtained through an adaptive calculation providing the cuts at a given poloidal plane and the strike surfaces. For the modeled shot, the experimental heat pattern and its time development are well described by the rotating unstable manifold, indicating the emergence of homoclinic lobes in a rotating frame due to the plasma instabilities.
\end {abstract}
\maketitle

\section {Introduction}\label{s1}

Modeling the interaction of the internal non-axisymmetric currents with the magnetic separatrix in single-null discharges is important to understand the influence of the plasma instabilities on the heat deposition in the divertor tiles. In ITER discharges a rotating non-axisymmetric divertor heat flux will cause periodic variation in the thermal loading of the target plates and cooling lines resulting in thermal and mechanical fatigue of these components leading to premature failures. This type of slowly rotating non-axisymmetric internal mode must be controlled to prevent damage to the divertor.

Experiments using internal 3D coils, such as those being installed in ITER, have been carried out in DIII-D to entrain these modes and prevent them from resulting in disruptions \cite{volpe2009}, but additional research is needed to better understand how to mitigate the thermal cycling of the divertors and this requires modeling the time-dependent evolution of the heat patterns caused by the plasma instabilities.

The concepts of invariant manifolds and magnetic footprints have proven useful to understand the interplay between magnetic asymmetries and heat flux patterns in poloidally diverted discharges~\cite{schmitz2011}. In single-null, axisymmetric, tokamak equilibria the plasma column is limited by a homoclinic magnetic separatrix. This magnetic surface is not resilient in three dimensions  and any departure from axisymmetry will lead to the splitting of the separatrix into two different surfaces named invariant manifolds~\cite{guckenheimer1983}. These manifolds intersect each other infinite times as they are stretched and folded in the neighborhood of the X-points \cite{guckenheimer1983, silva2002, roeder2003}. The existence of these non-axisymmetric surfaces has been demonstrated experimentally by measuring the heat deposition patterns in the divertor tiles \cite{evans2005}, and by tangential imaging of the X-point region \cite{kirk2012}, and is a subject of interest since the observation of helical footprints in tokamak discharges \cite{pompherey1992}.

The magnetic footprint can be obtained by following the magnetic field lines starting near the plasma edge until they intersect the material surfaces of the tokamak chamber. The location and connection lengths of the field lines are organized by the invariant manifolds of the magnetic saddle. These surfaces delimit the strike region and subdivide internally the laminar plots~\cite{wingen2009b}. Provided that the parallel transport is dominant for the modeled discharge the interior of the manifold corresponds in a first approximation with the heat pattern. In magnetically diverted configurations the field lines intersect the divertor targets in static spiral regions in agreement with the experimental heat distribution for L-mode plasmas~\cite{buttery1996, schmitz2011}, and with the particle flux distribution for H-mode plasmas~\cite{wingen2014}.

In more general situations the magnetic field changes in time and the field lines description becomes four-dimensional. In this scenario, the magnetic invariants and the strike patterns evolve in time and must be calculated for each moment. However, in the special case in which the magnetic perturbation rotates toroidally we can calculate the invariants in a rotating frame and then recover the temporal evolution in the laboratory frame. Experimental observations of heat deposition patterns consistent with the rigid rotation of an invariant manifold have been observed in \cite{evans2005}, where the heat pattern evolution was phase correlated with a slowly rotating or quasi-stationary tearing mode. However, a quantitative agreement between the heat pattern and the manifold calculation requires modeling the non-axisymmetric fields created by the plasma response. In this work, we model the magnetic field created by the internal asymmetric currents and show a quantitative agreement between the measured heat flux pattern at the divertor tiles and the area delimited by the unstable invariant manifold calculated for the DIII-D single-null discharge $\# 158826$.

For the discharge of interest the Electron Cyclotron Emission (ECE) signal reverses its phase at $q=2$, indicating the presence of a tearing island. Its time development is well correlated with the measured magnetic signals outside the plasma column. The slowly growing tearing mode is observed to rotate toroidally at an approximately constant rate. To model the non-axisymmetric field produced by the helical modes at $q=2$ we consider a set of helical filamentary currents resting at an internal magnetic surface of the EFIT axisymmetric equilibrium reconstruction of the shot \cite{lao1985}. This approach is similar to the one used to model the plasma response to applied Resonant Magnetic Perturbations (RPMs) in COMPASS-D \cite{buttery1996, cahyna2011}. However, in this situation, we do not prescribe the filaments using the X and O points induced by the RMPs, as this is inconvenient due to the proximity of the surface $q=2$ to the plasma edge. Conversely, we define our filaments in more internal surfaces to avoid undesired effects due to the discretization of the response currents, then we use a non-linear optimization technique to adjust the phases and currents of the filaments to reproduce the measured magnetic fluctuations at the magnetic probes in the mid-plane, low-field-side.

With the optimized model for the perturbation field we calculate the magnetic invariant manifolds in the rotating frame using a method similar to the described in~\cite{wingen2009a}, then we recover the expected time-dependent interior region delimited by the manifolds in the laboratory frame. The unstable manifold interior shows good agreement with the time-dependent heat flux profiles measured at the divertor plates in a fixed toroidal position, indicating the development of invariant manifolds in the rotating frame due to the internal non-stationary currents. The reasonable match of simplified approaches to model the time-dependent region affected by the heat flux is interesting for the development of effective fields that account for complicated self-consistent phenomena, which in principle requires the use of resistive Magnetohydrodynamics. 

This manuscript is organized as follows. In Section~\ref{s2} we present a summary of the experimental observations of the DIII-D discharge $\# 158826$. In Section~\ref{s3} we introduce the filamentary currents model and adjust its parameters to match the measured magnetic fluctuations and calculate the corresponding invariant manifolds for the equilibrium reconstruction. In Section~\ref{s5} we present our main result, namely the correspondence between the internal region delimited by the unstable manifold, and the measured time-dependent heat flux at the divertor tiles. In Section~\ref{s6} we present our conclusions and perspectives, and, finally, we have included some technical details of the manifold calculations in an Appendix section.

\section{Experimental observations}\label{s2}

The basic evidence of the interplay between the magnetic topology and the plasma edge, or, even, the definition of the plasma edge through the invariant manifolds, is the temporal correlation between the magnetic fluctuations at the magnetic probes and the evolution of the heat-flux pattern in the Plasma Facing Components (PFCs).

In this work, we consider the discharge $\# 158826$ of DIII-D where a small stationary $n=2$ field was applied. A non-rotating locked mode forms at $1625$ ms and remains stationary until $2015$ ms when it begins to rotate slowly at approximately $6.67$ Hz. During the non-rotating phase two heat flux peaks form at the outer strike point and remain fixed until the mode begins to rotate. 

\begin {figure}[h]
  \centering
  \includegraphics[width=0.45\textwidth]{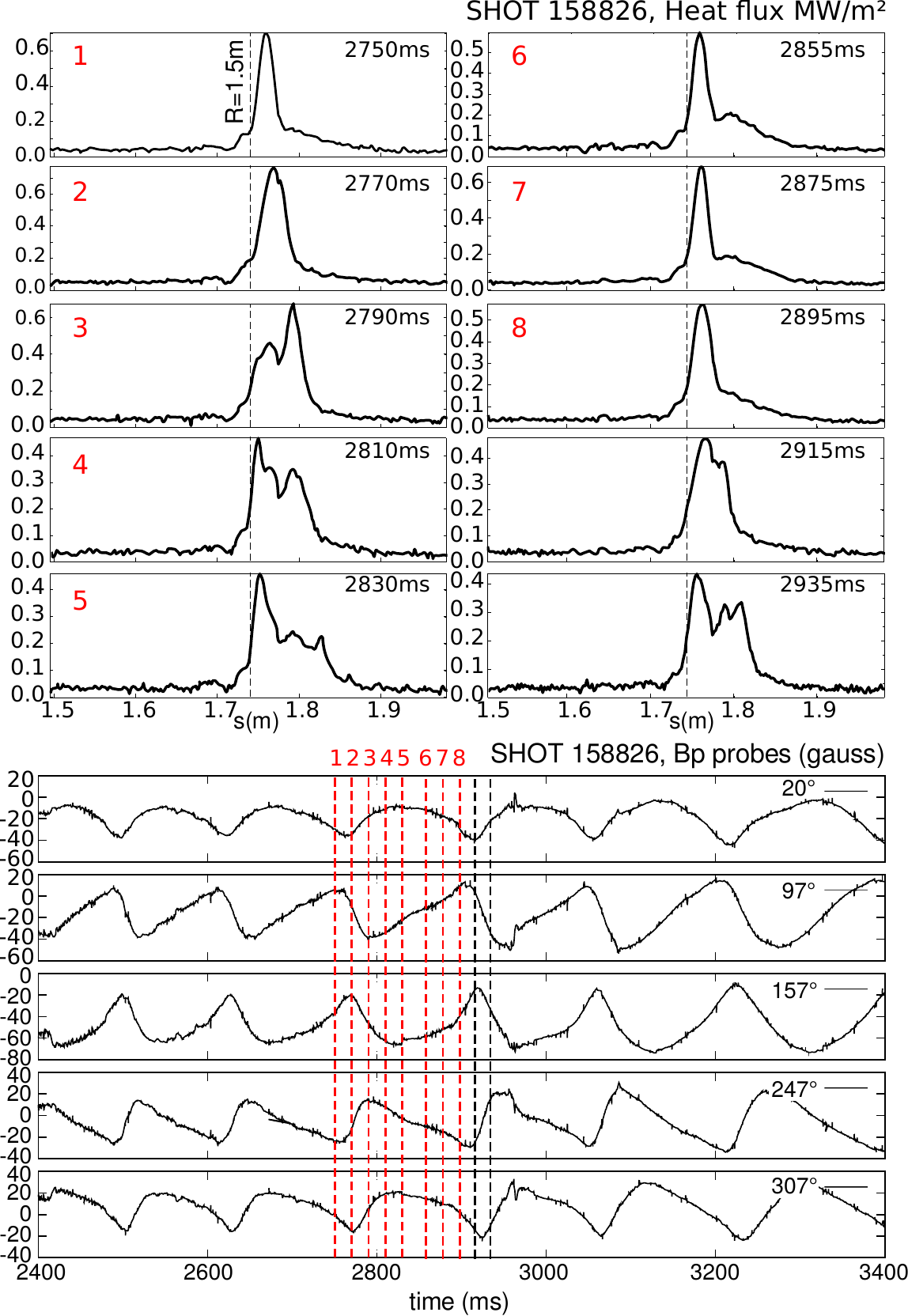}
  \caption{\label{f1} Heat flux profiles in the LFS strike point for discharge $\# 158826$ for ten time slices between $2750$ ms and $2935$ ms (top) and poloidal magnetic field fluctuations for various toroidal positions measured on the LFS mid-plane (bottom). The black dashed line in the heat flux profiles identify the strike point position obtained from the equilibrium reconstruction of the discharge, and the red dashed lines in the probe signals correspond to the times for the heat profile measurements during one period of the magnetic signals ($150$ ms). The coordinate $s$ in the heat flux profiles corresponds to the poloidal distance along the DIII-D wall, measured counterclockwise from the HFS mid-plane. For these profiles $s = R+0.239$ m, with $R$ the cylindrical radial coordinate.}
\end {figure}
	
In Fig.~\ref{f1} we show the poloidal magnetic field fluctuations at different toroidal positions in the Low Field Side (LFS) mid-plane and the corresponding measurements of the heat flux from an infrared (IR) camera pointing to the lower divertor region during the rotating stage.  As the magnetic signals complete one cycle, the heat flux profile completes one bifurcation cycle, where a single peak splits into two or three peaks that spread to the LFS and vanish.

This type of behavior is expected if the non-axisymmetric field created by an internal instability is changing periodically the magnetic topology, resulting in a periodic separation of the strike points measured by the IR camera \cite{evans2005}.

In this shot Electron Cyclotron Emission presents a phase inversion across the $q=2$ surface indicating the presence of a tearing island. Also, the modulation in the radial and poloidal magnetic fields measured at several toroidal positions for the same poloidal location are consistent with a $q=2$ tearing mode rotating at approximately $6.67$ Hz. The amplitude of the mode grows slowly, but it is modeled to be constant during the heat flux measurements. It is important to point out that although the phase differences between the signals are consistent with the rigid rotation of the mode, the waveforms captured at different toroidal locations present different shapes, indicating that the internal structure is not rotating rigidly, but may contain elements moving at different frequencies or even oscillating. However, these effects are not included in the model as they do not introduce information that can be resolved with the IR camera, also, the dominant mode provides substantial agreement with the available experimental data.

\section {Perturbation and separatrix}\label{s3}
\subsection {Helical filaments model for the perturbation}

In order to show a clear correlation between the magnetic signals and the divertor heat flux we will concentrate in modeling the time-dependent part of the magnetic field, which coincides experimentally with the heat flux becoming non-stationary. For this, we split the total magnetic field in the form
\begin{equation}\label{e1}
\vec B(R, z, \phi, t) = \vec B_0 (R,z) + \vec B_1 (R,z,\phi,t).
\end{equation}
Where $\vec B_0$ is the stationary, axisymmetric part of the magnetic field and $\vec B_1$ is the time-dependent non-axisymmetric part. The symmetric field $\vec B_0$ can be determined by solving the Grad-Shafranov equation while adjusting a set of plasma profiles consistent with the measured plasma parameters~\cite{lao1985}. For this part we use the EFIT equilibrium reconstruction of the discharge $\# 158826$ during the period of interest. For the time-dependent part we peform the following decomposition
\begin{equation}\label{e1b}
 \vec B_1 (R,z,\phi,t) = \vec B_{ext} (R,z,\phi) + \vec B_{resp} (R,z,\phi,t).
\end{equation}
Where $\vec B_{ext}$, represents the stationary, non-axisymmetric field created by external sources, like the I and C-coils of DIII-D, and $\vec B_{resp} (R,z,\phi,t)$ encompasses the time-dependent plasma response to the external perturbation and the field created by the internal instabilities like tearing or kink modes. As will be discussed at the end of this section, the ratio $|\vec B_{ext}|/|\vec B_{resp}|\lesssim 0.018$ in the separatrix region of discharge $\# 158826$. Consequently we approximate the total non-axisymmetric field with the response field alone
\begin{equation}\label{e1c}
 \vec B_1 (R,z,\phi,t) \approx \vec B_{resp} (R,z,\phi,t).
\end{equation}
Based on the previous section, the experimental phase shift between magnetic fluctuations at different probes suggests a dominant rigid rotation of the internal source currents, and a smaller contribution from non-rotating sources that reflect in the waveform difference between probes. In the following, we consider the influence of the dominant rotating component to obtain a simple three-dimensional rigid description that describes reasonably well the experimental heat patterns in Fig.~\ref{f1}.
To model this field we consider a non-axisymmetric current distribution inside the plasma, rotating toroidally at the measured dominant frequency $\nu_0 = 6.67Hz$, and creating the measured magnetic fluctuations outside the plasma. In the present situation, the rotation is quite slow, and we can approximate
\begin{equation}\label{e2}
 \vec B_{resp} (R,z,\phi,t) \approx \vec B_c(R,z,\phi-\Omega t),
\end{equation}
where $\vec B_c (R,z,\phi)$ is the stationary magnetic field created by the internal current as measured in a non-inertial rotating frame with angular velocity $\Omega = 2\pi\nu_0$. In a first approximation, the toroidal motion is considered to be uniform, so that the phase changes linearly with $t$.

For clarity, we will study the three-dimensional topology of the magnetic field in the rotating frame where the perturbation becomes stationary. Then we will recover the spatio-temporal dynamics in the laboratory frame to compare with the experimental observations. In a strict sense, $\vec B_0$ must be transformed to this non-inertial frame, but given the low frequency of rotation we use the field obtained through MHD equilibrium in the laboratory frame. To estimate the error of this assumption, notice that the fictitious force density $nM_iR_0\Omega^2$ is of the order $10^{-4} N/m^3$, while the laboratory magnetic forces $\vec j\times\vec B$ and pressure gradient $\nabla p$ are of the order $10^3N/m^3$. Consequently, no significant changes on $\vec j$ and $\vec B$ are expected in the rotating frame.

To model the perturbing field we consider a minimal set of internal helical filaments, carrying the non-axisymmetric currents inside the plasma and use the magnetic measurements at different angular positions to adjust simultaneously the currents and phases of all the filaments. This simple approach allow us to approximate the experimental magnetic field near the plasma edge without knowing the exact current sources, and without MHD calculations.
For a set of $N_f$ filaments with helicity $q_h$ on an \emph{arbitrary} magnetic surface, the non-axisymmetric part of the current density inside the plasma takes the form
\begin{equation}\label{e3}
  \vec j_1(\vec r, t) = \sum_{i=1}^{N_f}I_i\delta (\vec r -\vec r_h(\phi_i - \Omega t))\hat h,
\end{equation}
where $I_i$ is the current of the $i$'th filament, $\hat h$ is the current direction at $\vec r$ and $\vec r_h(\phi_0)$ is a helix with $q_h$ toroidal turns per poloidal turn. The desired helical path is obtained by appropriate scaling of the toroidal coordinate of a field line starting at the toroidal angle $\phi_0$ on the LFS mid-plane at the desired magnetic surface, so that the line helicity becomes $q_h$.

For the DIII-D discharge $\# 158826$ a tearing mode develops at the $q=2$ surface which is at approximately $8$ cm from the separatrix in the LFS. Placing the helical filaments in this location allows us to model the measured magnetic fluctuations at the probes. However, due to the small distance between the filaments and the separatrix, the discrete nature of the helical currents cause undesired topological effects, such as the deformation of the manifold around the filaments. Consequently, we do not use field aligned filaments at $q=2$, but consider filaments with $q_h=2$ in a more internal surface at $40\mbox{cm}$ from the separatrix in the LFS mid-plane. The number of filaments is chosen by requiring a maximum drop in the fitting error for an small increase in the number of filaments. \emph{Three} filaments gives a large improvement relative to \emph{two} or \emph{one} filament, while more than \emph{three} gives a poor relative improvement.

Our approximations are only applicable for studying the perturbed field lines outside the magnetic surfaces where the instability is formed, and are not intended to model the internal magnetic topology self-consistently, this is why we locate the perturbation sources arbitrarily, whenever we can reproduce the measured magnetic fluctuations. The location of the source currents do affect the internal structure of the invariant manifolds, but their more explicit features like number of lobes, lobe width, and the lobe radial excursion, are mostly determined by the dominant mode numbers and their relative intensities, and these figures can be conserved when changing the location of the sources. For the modeled filaments, the ratio between poloidal and radial magnetic fluctuations created at the separatrix depends weakly on the chosen magnetic surface when the distance between the separatrix and the filaments is above $10\mbox{cm}$.

\begin{figure}[h]
  \centering
  \includegraphics[width=0.4\textwidth]{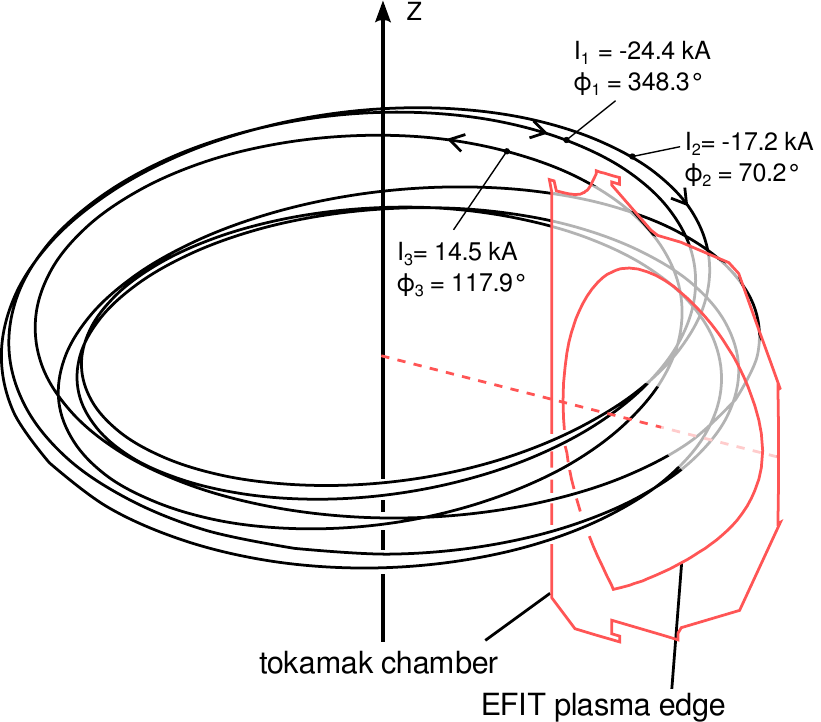}
  \caption{\label{f2} The error functional is well minimized by three internal helical filaments with $q_h=2$ at $40$cm from the separatrix in the LFS mid-plane, rotating toroidally at the measured dominant frequency 6.67Hz.}
\end{figure}

The current in the filaments $\{I_1, I_2, I_3\}$ and their toroidal phases $\{\phi_1, \phi_2, \phi_3\}$ are determined simultaneously using a Levenberg-Marquardt routine \cite{marquardt1963} to minimize the corresponding error functional
\begin{equation}
  \epsilon (\{I_i,\phi_i\}) = \sum_{n=1}^N\sum_{m=1}^M[\tilde B^{(m)}_z(t_n) - B_1(\{I_i,\phi_i\},\Omega t_n - \tilde\phi_m)]^2.
\end{equation}
Where $\tilde B^{(m)}_z(t_n)$ is the poloidal field fluctuation measured by the $m'$th magnetic probe located in $\tilde\phi_m$ for the time slice $t_n$, and $\{N, M\}$ are the number of time slices and magnetic probes respectively. The simulated magnetic perturbation $B_1(\{I_i,\phi_i\},\phi)$ is calculated for the corresponding angles of each probe in the rotating frame, where $\Omega = 2\pi\times 6.67$ Hz is the measured toroidal rotation frequency, and the adjusted time interval contains the heat measurements in Fig.~\ref{f1}. When provided a reasonable initial guess for the parameters, the routine converges to a minimum value of the error, where $\{I_1,I_2,I_3\} = \{24.4, -17.2, -14.5\}$ kA and $\{\phi_1,\phi_2,\phi_3\} = \{348.3\textdegree, 70.2\textdegree, 117.9\textdegree\}$ (Fig.~\ref{f2}).

\begin{figure}[h]
  \centering
  \includegraphics[width=0.45\textwidth]{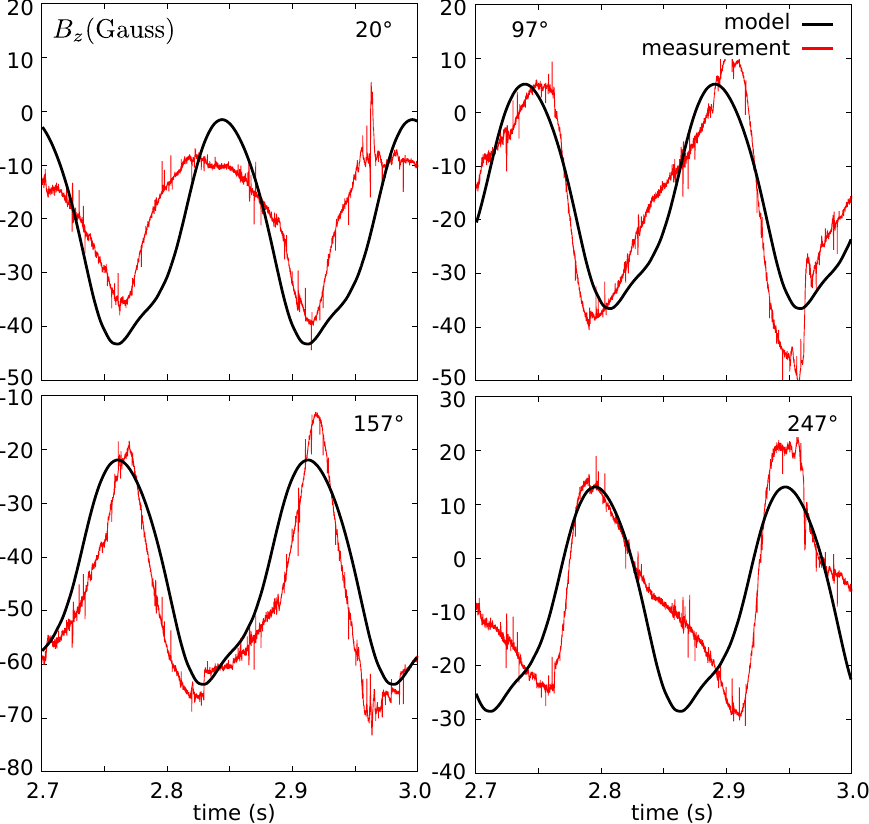}
  \caption{\label{f3} Comparison between the magnetic probe signals presented in Fig.~\ref{f1} at various toroidal locations (red), and the vacuum field created by the modeled rotating internal filaments at the probe locations (black). A reasonable match between maxima and minima is generally observed, while the waveform shape is not necessarily reproduced at some locations due to the approximation of rigid rotation.}
\end{figure}

In Fig.~\ref{f3} we show the comparison between the poloidal magnetic field measured by four magnetic probes at $(R,z)=(2.4, 0.0)$ m in different toroidal positions, and the modeled field value from the helical currents using a Biot-Savart calculator. For the modeled portion of the shot between $2700$ and $3000$ ms the filament currents and their relative phases were kept constant in the rotating frame. Between $2400$ and $2900$ ms, the magnetic fluctuations have an approximately constant amplitude and frequency, but at the end of the heat flux measurements, after $2900$ ms, the magnetic fluctuations are slightly increased.

After the parameter optimization, the filamentary currents account for the amplitude, phase, frequency and general shape of the magnetic oscillations in most of the magnetic probes signals, with some variations caused by small departures from a rigid rotation of the currents comprising the plasma response (Fig.~\ref{f3}).

The departure from rigid rotation of the internal currents may be due to the magnetic torque created by stationary magnetic fields produced internally and externally, but modeling the interaction of the stationary field with the model filaments is outside the scope of this work. Also, a simple addition of the static fields to the response field without interaction with the filaments will not affect the modeling of the magnetic signals because stationary fields only adds a bias to the simulation of the measurement, without affecting the waveform. However, the addition of a stationary fields for the manifold calculations can reshape the lobes in time causing measurable effects in the heat flux patterns, but this effect will only be observed if the stationary field plays a significant role in the definition of the non-axisymmetric field in the separatrix region.

In the discharge of interest the I and C-coils where operating during the heat measurement, causing possible departure from the rigid rotation in the manifolds. To estimate the importance of this external field $\vec B_{ext}$, it was calculated in a uniform grid along the unperturbed separatrix and compared to the field created by the internal filaments $\vec B_{resp}$.
\begin{figure}[h]
  \centering
  \includegraphics[width=0.4\textwidth]{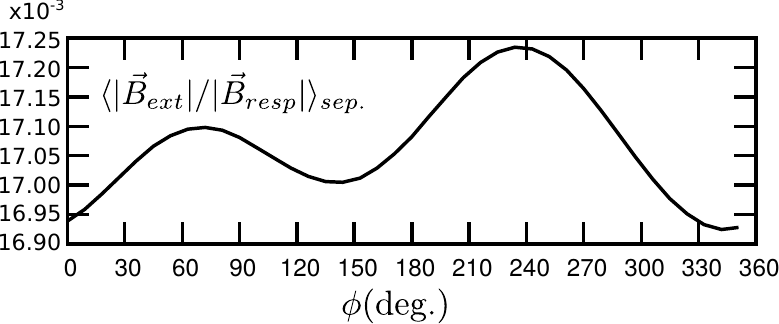}
  \caption{\label{f3b} Magnetic surface mean of the ratio between the fields created by the external stationary currents and the internal helical currents at the magnetic separatrix for the DIII-D shot $\#158826$ as a function of the rotation phase of the internal filaments.}
\end{figure}
In Fig.~\ref{f3b} we show the mean of the ratio $|\vec B_{ext}|/|\vec B_{resp}|$ at the magnetic separatrix as a function of the phase difference between helical filaments and the static array of coils. Clearly, the magnetic field from stationary sources would only modify the non-axisymmetric field in the separatrix region by less than $2\%$, while adding unnecessary complexity to the analysis of the manifolds in space and time. In the following, we disregard the stationary fields and consider the rigid rotating filaments as the only source of non-axisymmetry, which allow us to perform a simplified description of the non-axisymmetric field in the rotating frame and calculate the corresponding non-axisymmetric separatrix that expectedly delimits the time-dependent region of increased heat flux at the divertor targets.

\subsection{Calculation of the magnetic separatrix}

In axisymmetric situations the magnetic field lines inside the plasma span a continuous set of nested toroidal magnetic surfaces. In diverted magnetic configurations these toroidal surfaces exist inside a magnetic separatrix containing one or more X-points (magnetic saddles). More precisely, the separatrix is defined by the field lines that converge to the X-points when followed in any direction, and in most situations the separatrix is spanned by the field lines converging to a single X-point.

In the axisymmetric case, the combined X-points at every poloidal section define a circle $(R,z) = (R_s, z_s), 0\leq\phi<2\pi$, where the poloidal field vanishes and the Jacobian of $\vec B_0$ has real eigenvalues. As the magnetic field is purely toroidal, a field line starting at $(R_s,z_s)$ is a closed circle passing through every X-point. In non-axisymmetric situations we can say, equivalently, that the X-points are defined by a field line closing after one toroidal cycle. In the following we will refer to this field line as an Unstable Periodic Orbit (UPO)\cite{guckenheimer1983}.

With this we can implement the definition of a separatrix for non-axisymmetric situations with an alteration. Now, the surface is defined by the field lines converging to the UPO when followed in one direction. This causes the existence of two different surfaces or invariant manifolds defining the confined region. The \emph{unstable manifold} is spanned by field lines converging to the UPO when followed opposite to the field direction, and, conversely, the \emph{stable manifold} is spanned by the field lines converging to the UPO when followed in the direction of the field. Both surfaces are infinite and non-self-intersecting, but can intersect each other in very intricate patterns.  The set of orbits converging to the UPO when followed in any direction lie on the intersections of the stable and unstable manifolds, referred to as homoclinic points, and they do not span a continuous surface, i.e. there are gaps between these orbits and they can not be the generalization of the axisymmetric separatrix.

In non-axisymmetric situations the UPO is a global structure and can not be identified using only local information. This happens because the toroidal angle $\phi$ is no longer a cyclic coordinate, i.e. the vanishing of the poloidal magnetic field can not be used to determine the UPO. To locate this orbit we need to identify the fixed point of the Poincaré Map $\tilde{P}$, at a given plane transverse to the magnetic field (see Appendix). Once identified this point it can be used to build the UPO by integrating the magnetic field starting at $\tilde{P}$ until it closes after one toroidal cycle.

In the neighborhood of the fixed point $\tilde{P}$, at the transverse plane, the invariant manifolds can be approximated using polynomials that adjust the behavior of neighboring field lines. Then, small segments of the manifold cuts can be mapped forward or backwards to build the rest of the manifold cuts far from the UPO. To perform this calculation, the {\sc magman} routine was developed (acronym for {\sc mag}netic {\sc man}ifolds). {\sc Magman} provides an adaptive calculation of any manifold cut by introducing new orbits \emph{on the fly} avoiding redundant calculations and spurious points, resulting in a well spaced sequence of linkable points that represent the invariant manifold intersection with the transverse plane (see Appendix).

Using the methods described in the Appendix, we were able to identify the intersection of the UPO with the plane $\phi=0$ for the equilibrium perturbed by the three filamentary currents in Fig. 2. The fixed point was located at $(R_s, z_s) = (1.3931, -1.1023)$ m, at approximately $3.7$ cm from the unperturbed saddle $(R_s^0,z_s^0) = (1.3816, -1.0672)$ m. Then we calculated the intersection of the invariant manifolds with the poloidal plane at $\phi=0$ using the adaptive advection of elementary segments implemented in {\sc magman}.

\begin{figure}[h]
  \centering
  \includegraphics[width=0.45\textwidth]{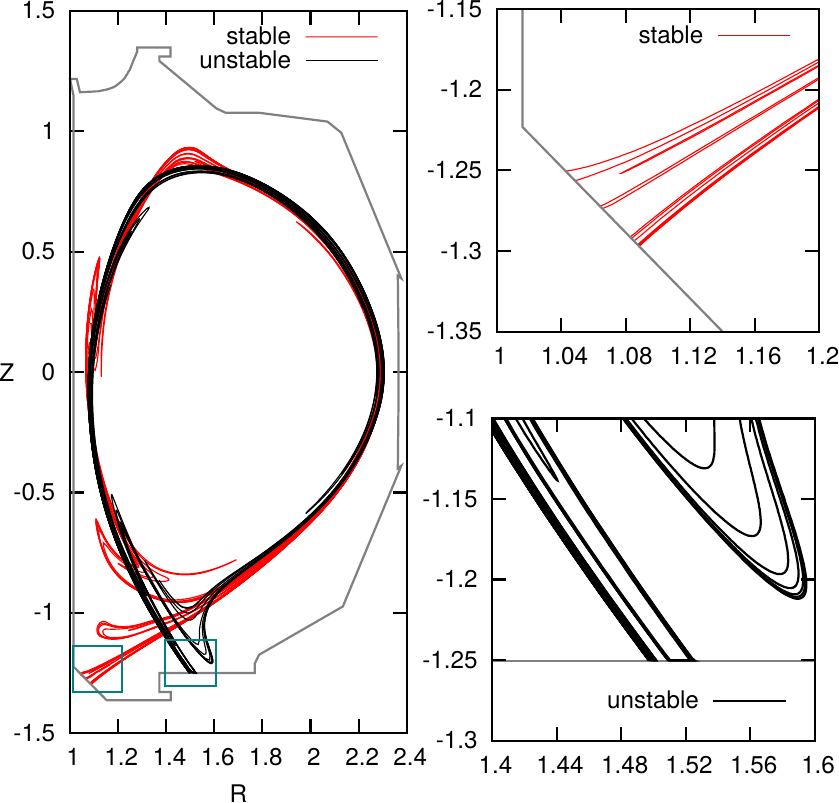} 
  \caption {\label{f4} Cut of the invariant manifolds in the plane $\phi=0$, for discharge $\# 158826$ subjected to the non-axisymmetric field created by three helical filaments at the surface $q=1$. The detail of the manifold intersection with the chamber reveals the layered interior of the magnetic footprints.}
\end{figure}

In Fig.~\ref{f4}, we show the cut of the invariant manifolds at $\phi = 0$ in the rotating frame, where the helical filaments are stationary. The invariant manifolds develop the characteristic \emph{homoclinic lobes} \cite{guckenheimer1983} created by the magnetic perturbation on the invariant manifolds as they get stretched and folded to preserve the toroidal magnetic flux. The multiple intersections of the invariant manifolds lead to the proliferation of UPOs of higher periods by the \emph{horseshoe mechanism} \cite{smale1967} which drives the chaotic behavior of the field lines at the plasma edge.

 \section {Results and discussion}\label{s5}

As we move toroidally in the rotating frame, (or as the time advances in the laboratory frame), the lobes move towards the UPO, and get stretched and folded. During the stretching, the invariant manifold crosses the tokamak wall in several regions that rotate toroidally in the laboratory frame.

These crossings can be calculated with {\sc magman} in the same way that the continuous poloidal section was calculated. The manifold cut at the tokamak chamber delimits the magnetic footprints or escape patterns and defines the internal boundaries through which there are abrupt changes in the connection lengths in a laminar plot~\cite{wingen2009b}.

To compare the {\sc magman} manifolds in the rotating frame with the heat deposition patterns measured at a fixed location in the laboratory we need to transform the times of the heat measurements to the corresponding toroidal phases of the rotating frame. This is achieved by
\begin{equation}\label{e5}
  \phi'_i =\phi_c - \Omega(t_i-t_0),
\end{equation}

where $t_i$ is the time of the $i$’th measurement, $\phi_c$ is the toroidal phase of the IR camera, and $\Omega = 2\pi\times 6.67$ Hz was determined from the waveforms captured by the magnetic probes.
 
\begin{figure}[h]
  \centering
  \includegraphics[width=0.45\textwidth]{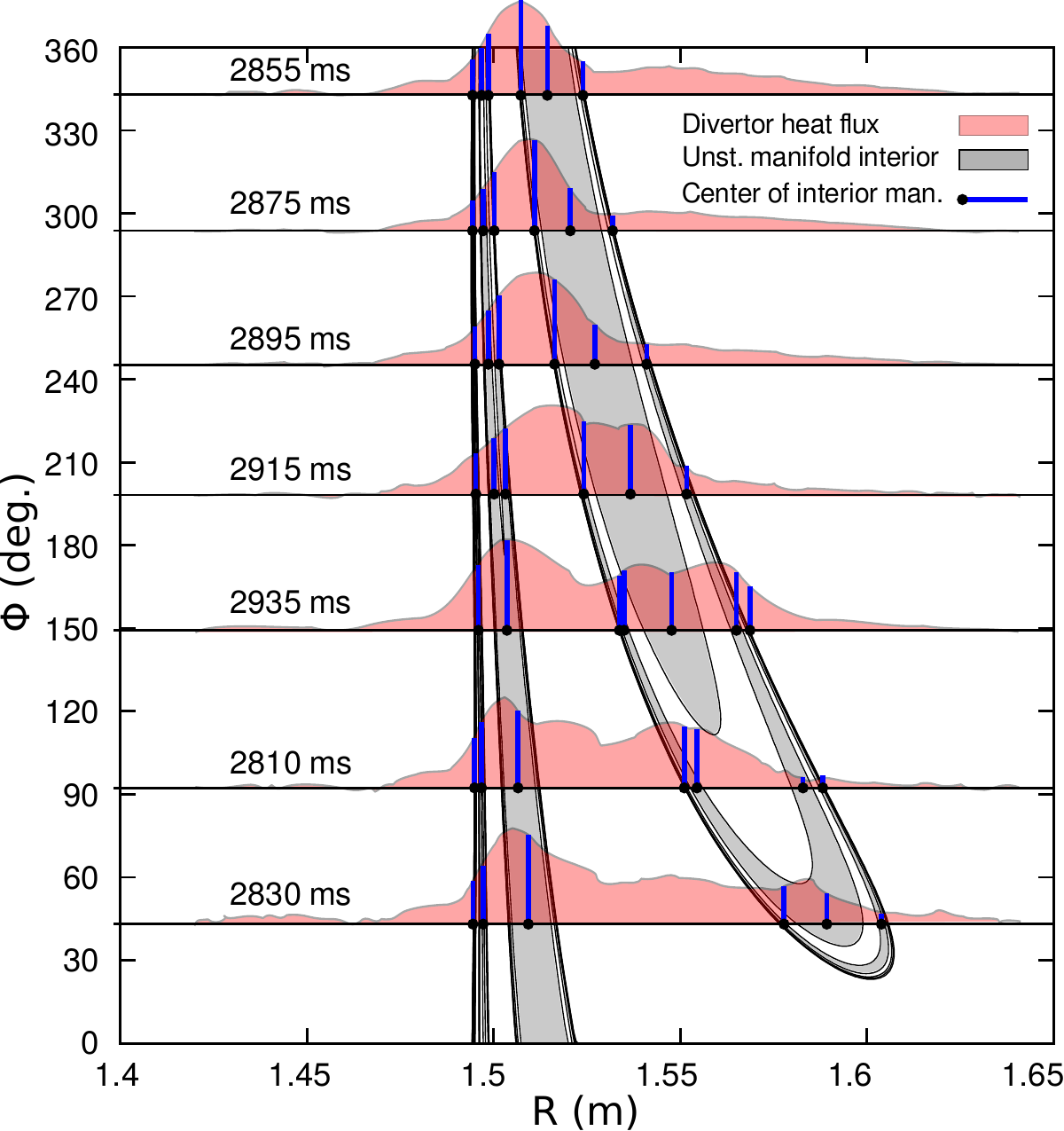}
  \caption{\label{f5} The interior of the unstable invariant manifold intersecting the DIII-D divertor plates in discharge $\# 158826$ in a rotating frame with toroidal frequency of $6.67$ Hz (gray region), and an overlay of the experimental heat flux measurements for the corresponding toroidal phases of the rotating frame (profiles filled in light-red). The heat flux peaks are well correlated with the center of the manifold interior (blue lines) for most phases, and the secondary heat flux peak disappears where the lobes get too close radially to be resolved by the IR camera. The heat pattern evolves periodically, as expected if the manifold rotates rigidly.}
\end{figure}

In Fig.~\ref{f5} we depict the intersection of the unstable manifold with the chamber wall for the LFS strike region in the rotating frame. Overlapped on the manifold, we depict the experimental heat flux measurements introduced in Fig.~\ref{f1} at the phases in the rotating frame corresponding to the time of each heat flux profile in the laboratory frame.

Notice that the heat flow peaks match reasonably well with the manifold interiors, and the observed peak splitting agrees well with the multiple interiors for each toroidal phase of the rotating frame. As discussed before, the invariant manifolds delimit the volume of confined field lines in non-axisymmetric situations. Consequently, the field lines inside the manifold can access the plasma interior driving the hot plasma by parallel transport to the Plasma Facing Components (PFC), while the field lines outside the manifold can not access the plasma and will only drive particles that leave the manifold interior by cross-field transport, which is small compared to the parallel transport.

It is interesting to notice that, for $240\textdegree<\phi<360\textdegree$, the camera sees a single heat peak moving from $1.520$ m to $1.525$ m, which is well aligned with the clustered lobes, but as the lobes get separated the peak splits and the main peak moves to the HFS following another piece of the manifold with a smaller radial extension. Clearly, the heat load is not aligned with the axisymmetric strike line from EFIT at $R\approx 1.5$ m but follows the wider lobes. This behavior explains the small misalignment between the main flux peak and the strike point position from the axisymmetric reconstruction in Fig.~\ref{f1}. This measurement is consistent with the invariant manifold delimiting the region of enhanced heat flux in non-axisymmetric discharges and indicates that the heat flux may be concentrated in the lobes instead of the region where the manifold gets compressed near the unperturbed axisymmetric strike line.

An important point is that the calculation of the manifold interior is not intended to describe the heat flux profile, but to identify the regions of the chamber affected by parallel transport. Heat flux profile features as intensity, width and peak locations require transport simulations that include particle drifts~\cite{wingen2014}, but it is interesting to notice that the plasma edge calculation through invariant manifolds delimits well the radial extension of the heat load and approximates well the locations of the main peaks.

Another approach involves modeling the equilibrium with fully three-dimensional solvers using energy minimization procedures. However, most tractable procedures extremize the plasma potential energy by performing ideal variations of the magnetic field, precluding topological changes at the resonant surfaces and the plasma edge. This preserves numerically the nested topology and prevents the numerical simulations to reproduce the experimentally observed separatrix splitting~\cite{hudson2010}. On the other side, resistive MHD simulations of the tearing mode can be used to calculate the non-axisymmetric fields created in the separatrix, but these calculations are numerically more involved and require the definition of a considerable set of model parameters. The presented treatment can be considered as an effective field modeling based on magnetic measurements that intends to pack a large number of complex effects in a \emph{simple} non-axisymmetric field.

The identification of the invariant manifolds with the plasma edge brings interesting possibilities for the interpretation and modeling of plasma structures near the plasma-vacuum interface and the evolution of the heat loads on the chamber components. These structures may emerge from parallel transport along the field lines resulting from the overlapping of dynamic and stationary magnetic fields in the presence of a magnetic saddle. This general situation may be relevant during the ELM suppression by Resonant Magnetic Perturbations where a non-stationary magnetic perturbation caused by evolving currents in the plasma is combined with an externally applied magnetic perturbation. Also, the so-called filamentary structures may be a manifestation of time-dependent helical features developing at the invariant manifolds~\cite{eich2003}, but this may require more detailed modeling of the non-axisymmetric source currents.

\section{Conclusions}\label{s6}

In this work we have shown that the measured heat flux to the divertor plates of a single-null discharge evolves in agreement with the interior of a rotating invariant manifold calculated for a minimalist model of the plasma response field created by a rotating tearing instability. The non-axisymmetric field was calculated with a filamentary model of the source currents consisting of three helical wires with helicity $q_h=2$ located at $40$cm from the separatrix in the mid-plane. The model parameters were adjusted to reproduce the measured magnetic signals at several probes outside the plasma. A more sophisticated treatment involving multiple helicities, several surfaces and more filamentary currents can provide more details of the heat patterns and the topological changes experienced by the manifolds, but in terms of experimental observations a simple model with reasonable properties gives a good quantitative match, suggesting that the extension and overall shape of the heat pattern is mainly controlled by the size and wave numbers of the magnetic perturbations at the plasma edge. The study of the interplay between the magnetic topology and the heat flow provides a relevant insight on the nature of the helical structures developing at the plasma edge. Whenever the parallel transport is dominant we can describe important features of real discharges using simple effective models like the one presented here.

\section* {Acknowledgments}
This work has been supported by Brazilian scientific agencies CAPES, CNPq and the S\~ao Paulo Research Foundation (FAPESP) under grants 2011/19269-11, 2012/18073-1 and 2014/03899-7, and by the U.S. Department of Energy, Office of Science, Office of Fusion Energy Science under DOE awards: DE-FC02-04ER54698, DE-SC0012706 and DE-FG02-05ER54809. The authors wish to thank Dr. Michael A. Makowski for his help on interpreting the heat flux data.

\section {Appendix}
\subsection{Unstable Periodic Orbits and Invariant Manifolds}

\begin{figure}[h]
  \centering
  \includegraphics[width=0.45\textwidth]{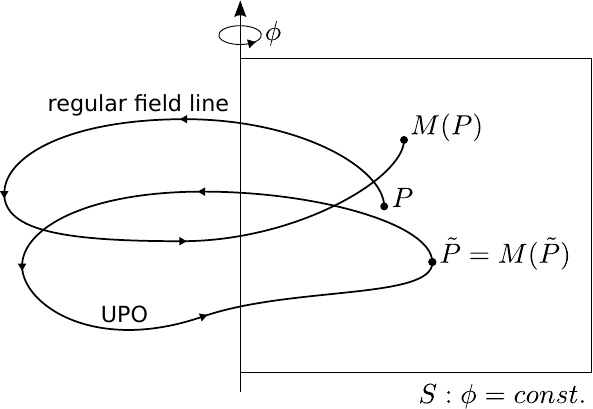} 
  \caption{\label{f6} Intersections of a regular field line and the UPO with a Poincaré plane.}
\end{figure}

To localize the Unstable Periodic Orbit (UPO), or saddle orbit, let us introduce the Poincaré map $M:S\rightarrow S$, where $S$ is a given surface, for instance $\phi=\mbox{const.}$, which is always transverse to the magnetic field, and $M(P)$ is the position of a field-line starting at $P\in S$ after one toroidal transit in the direction of the field (Fig.~\ref{f6}). The inverse map $M^{-1}$ is obtained when we follow the line in opposite direction to the field. Notice that, $M(M{-1}(P)) = M^{-1}(M(P)) = P$.

The explicit form of $M$ is unknown and we can only calculate it numerically through the integration of the field lines. The field line tracing is performed with an order five adaptive Runge-Kutta integrator \cite{cash1990}. The error estimate was adjusted to $10^{-12}$ m per step, so that, cumulative errors are expected to be small.

Formally, we can look for the fixed point $\tilde P$ satisfying $M(\tilde P) = \tilde P$, which belongs to the UPO (Fig.~\ref{f6}). This can be accomplished numerically using a Levenberg-Marquardt algorithm \cite{marquardt1963} with a tracking stage or the Broyden's method \cite{broyden1965}. Both methods involve the approximation of the Jacobian matrix of $M$ using finite differences.  Provided a good initial guess of $\tilde P$ the method converges in few iterations, usually, less than $10$. For a perturbed single-null discharge the unperturbed saddle becomes good initial guess for the UPO.

\begin{figure}[h]
  \centering
  \includegraphics[width = 0.45\textwidth]{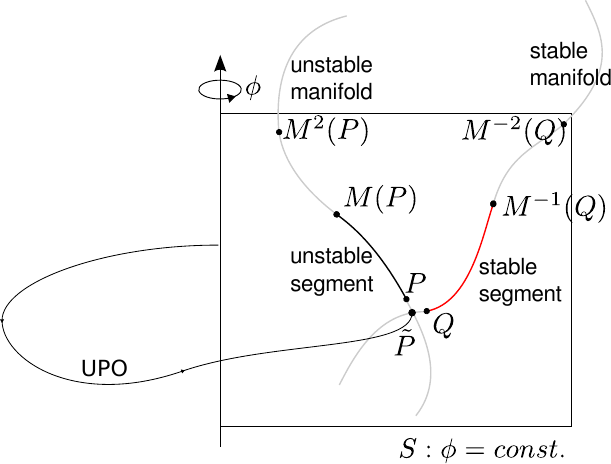}
  \caption{\label{f7} The invariant manifolds are calculated by applying the Poincaré map to small segments of the invariants close to the UPO. The local invariants are obtained by fitting the mapped points from a neighborhood of $\tilde{P}$.}
\end{figure}

In the neighborhood of the fixed point $\tilde P\in S$ the invariant manifolds dominate the geometry of the field lines (Fig.~\ref{f7}). A localized set of points around $\tilde P$ will be stretched along the unstable manifold as we apply the Poincaré map $M$ repeatedly, and the same set of points will be stretched along the stable manifold upon repeated applications of $M^{-1}$.
To only visualize the manifold cut it is sufficient to initialize a large number of points around $\tilde P$, and apply repeatedly the map or its inverse respectively. However, this method results in a scattered representation of the manifold, which can not be optimized in a controlled fashion and can not be generalized to represent the manifold in three-dimensions, which may be desirable for comparison with tangential imaging of the plasma edge. In this work we use the local behavior of the field lines around $\tilde P$ to calculate a local representation of the manifolds by adjusting a set of stretched points with a polynomial of order $5$ or $7$ containing the fixed point for both the stable and unstable manifolds. Then we use an adaptive algorithm to build the rest of the manifold in the transverse surface $S$.

\begin{figure}[h]
  \centering
  \includegraphics[width=0.45\textwidth]{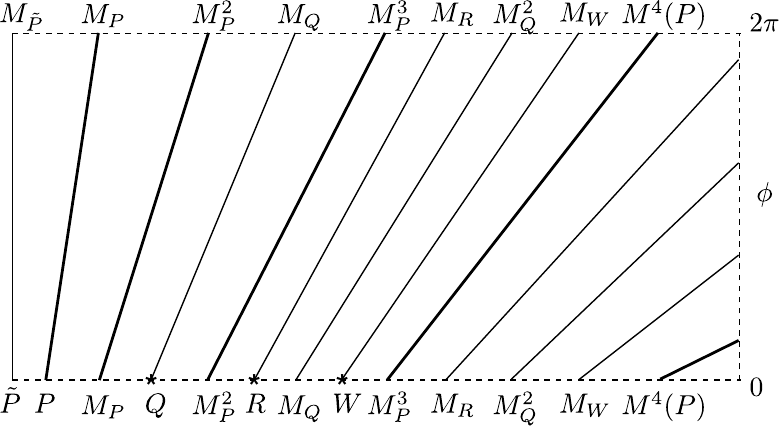}
  \caption{\label{f8} Representation of the optimization process. Each line represents a field line on the invariant manifold. At the top and bottom we have the intersections with the Poincar\'e plane $\phi=0$, where $M^N_P = M^N(P)$. After the first cycle, the distance between $M_P$ and $M^2_P$ was larger than the critical distance $d_c$, then a new orbit is created at $Q$. The refinement continues, by measuring the distance between $M^2_P$ and $M_Q$ that is also larger than $d_c$, originating $R$ and so on. The starting points of the new orbits are calculated in the cutting plane using Lagrange polynomials passing through the neighboring points.}
\end{figure}

For this task we use the {\sc magman} routine, which consists in applying the Poincaré map repeatedly to a local segment of the manifold around $\tilde{P}$ (Fig.~\ref{f7}). The spatial resolution of the manifold is preserved by means of a refinement procedure that enables the introduction of new field lines where they are needed (Fig.~\ref{f8}), avoiding the introduction of new orbits very close to $\tilde{P}$ and preventing the clustering of points in other regions of the manifold. This method uses the points describing the manifold segment in the previous iteration to calculate new initial conditions at the places where the distance between consecutive points becomes larger that some critical distance $d_c.$ This critical distance was estimated by requiring that the lobes were well resolved. For this, we estimated the characteristic lobe size and required it to be one order of magnitude larger that $d_c$.

The new points are calculated with Lagrange polynomials at the cutting surfaces to insert the required new orbits smoothly within the sequence of existing ones. The {\sc magman} routine is similar to the {\sc mafot} about the propagation of local segments \cite{wingen2009a}, but differs in the dynamic refinement of the segment. {\sc magman} also allows us to continue the manifold along material objects, like the divertor tiles and the vacuum chamber without extra calculations or resolution loses, also, it returns an organized sequence of points that can be connected with segments to represent the manifold.

\begin{figure*}[t]
  \centering
  \includegraphics[width=0.9\textwidth]{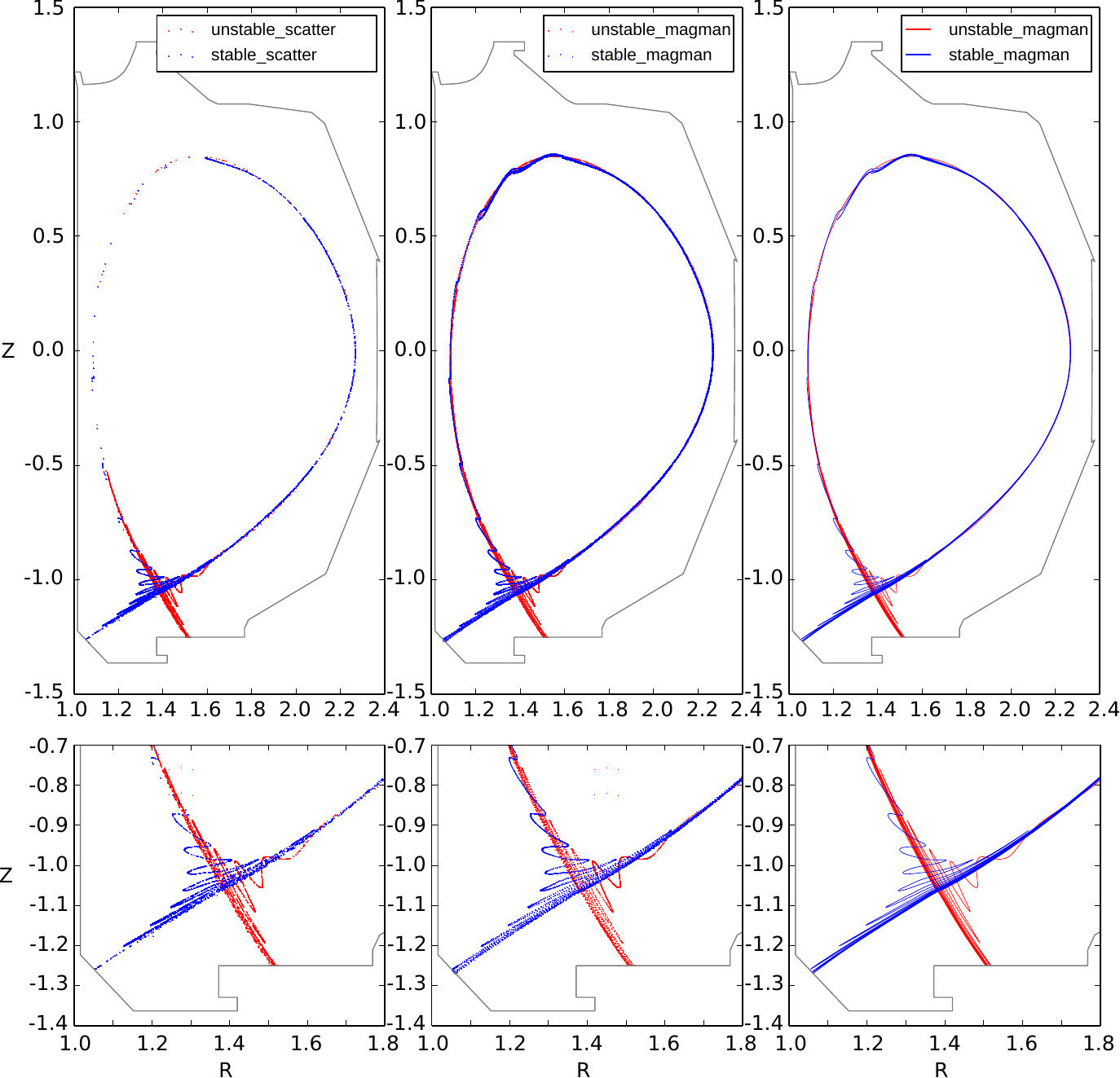}
  \caption{\label{f9} The left column shows the scatter-plot of a set of $7500$ random initial conditions close to the UPO mapped for 9 toroidal cycles. The center column shows the scatter plot of the {\sc magman} adaptive calculation, with the same number of points. Clearly, the {\sc magman} points are better distributed along the manifold and reach farther regions. The third column shows the smooth line traced along the organized {\sc magman} points. As the random sampling does not provide organized points, they can not be joined smoothly to represent the manifold.}
\end{figure*}

To illustrate the importance of having an organized and well distributed sequence of points we calculate the invariant manifolds of the EFIT equilibrium reconstruction for the shot $\# 158826$, when subjected to an external $n=3$ perturbation. For comparison, we represent the manifolds by mapping a set of $7500$ random initial conditions close to the UPO for $9$ toroidal cycles. In Fig.~\ref{f9} we compare the resulting scatter plots from the random sampling with those from {\sc magman}, which in addition are organized and can be joined smoothly.

The comparison is performed so that the number of points representing the manifold in the random and adaptive calculations is the same. For the presented results, the {\sc magman} calculation requires only $10\%$ of the mappings required by the random sampling method and involves only $5\%$ of its integration steps and computation time, this happens because the lobes of the manifold develop in regions where the adaptive step integrator performs larger steps. If the comparison is performed so that we obtain the same quality representation with both methods the number of random initial conditions must be tripled, and the number of required mappings for the random method will be $30$ times the required by {\sc magman}. Consequently, the computation time would be about $60$ times the required by {\sc magman} for a plot with the same quality.

\section*{References}
\bibliography{references}
\bibliographystyle{iopart-num}

\end {document}